\documentclass{jfm}
\usepackage{graphicx,psfrag}
\usepackage{caption} 
\usepackage{graphicx}
\usepackage{subcaption}
\usepackage{natbib}
\usepackage{float}
\usepackage{rotating}
\usepackage{amsmath}
\usepackage{latexsym}
\usepackage{amssymb}
\usepackage{mathrsfs}
\usepackage{psfrag}
\usepackage{layout}
\usepackage{bm}
\usepackage{url}
\usepackage{epstopdf}
\usepackage{wrapfig}
\usepackage{tabularx}
\usepackage{fancyhdr}
\usepackage{color}
\usepackage[makeroom]{cancel}

\ifCUPmtlplainloaded \else
  \checkfont{eurm10}
  \iffontfound
    \IfFileExists{upmath.sty}
      {\typeout{^^JFound AMS Euler Roman fonts on the system,
                   using the 'upmath' package.^^J}%
       \usepackage{upmath}}
      {\typeout{^^JFound AMS Euler Roman fonts on the system, but you
                   dont seem to have the}%
       \typeout{'upmath' package installed. JFM.cls can take advantage
                 of these fonts,^^Jif you use 'upmath' package.^^J}%
      }
  \else
  \fi
\fi

\ifCUPmtlplainloaded \else
  \checkfont{msam10}
  \iffontfound
    \IfFileExists{amssymb.sty}
      {\typeout{^^JFound AMS Symbol fonts on the system, using the
                'amssymb' package.^^J}%
       \usepackage{amssymb}%
       \let\le=\leqslant  \let\leq=\leqslant
         \let\geq=\geqslant
      }{}
  \fi
\fi

\ifCUPmtlplainloaded \else
  \IfFileExists{amsbsy.sty}
    {\typeout{^^JFound the 'amsbsy' package on the system, using it.^^J}%
     \usepackage{amsbsy}}
    {\providecommand\boldsymbol[1]{\mbox{\boldmath $##1$}}}
\fi

\newcommand{\Rel}{$Re_\lambda$}

\newcommand{\mi}{\mathrm{i}}

\DeclareMathAlphabet\mathsfbi{OT1}{cmss}{m}{sl}

\providecommand\bcdot{\boldsymbol{\cdot}}
\newcommand{\bs}{\boldsymbol}

\usepackage{tikz}

\title[]{Clustering of Rapidly Settling, Low-Inertia 
Particle Pairs in Isotropic Turbulence. \\ II. Comparison of Theory and DNS}

\author[Sarma L. Rani, Rohit Dhariwal, and Donald L. Koch]%
{S\ls A\ls R\ls M\ls A\ns L.\ns R\ls A\ls N\ls I$^1$%
  \thanks{Email address for correspondence: sarma.rani@uah.edu},
 R\ls O\ls H\ls I\ls T\ns D\ls H\ls A\ls R\ls I\ls W\ls A\ls L$^2$, \and 
\break
D\ls O\ls N\ls A\ls L\ls D\ns L.\ns K\ls O\ls C\ls H$^3$}

\affiliation{$^1$Department of Mechanical and Aerospace Engineering, University of Alabama in Huntsville,
Huntsville, Alabama 35899, U.S.A.\\[\affilskip]
$^2$Department   of Civil and Environmental Engineering, Duke 
University, Durham, North Carolina 27708, U.S.A.\\[\affilskip]
$^3$School of Chemical and Biomolecular Engineering, Cornell University, Ithaca, New York 14853, U.S.A.}

\begin{document}

\maketitle
\begin{abstract}

Part I of this study presented a stochastic theory 
for the clustering of monodisperse, rapidly settling, low-Stokes-number  
particle pairs in homogeneous isotropic turbulence.    
The theory involved the development of closure approximations for the drift and diffusion 
fluxes in the probability density function (PDF) equation for pair relative positions.  
%
In this Part II paper, the theory is quantitatively analyzed 
by comparing its predictions of particle clustering with  
data from direct numerical simulations (DNS) of isotropic 
turbulence containing particles settling under gravity.  
DNS were performed at a Taylor micro-scale Reynolds number $Re_\lambda = 77.76$ 
for three Froude numbers $Fr = \infty,~ 0.052,~ 0.006$.  
The Froude number $Fr$ is defined as the ratio of the Kolmogorov scale of 
acceleration and the magnitude of gravitational acceleration.  
Thus, $Fr = \infty$ corresponds to zero gravity, and $Fr = 0.006$ to the highest 
magnitude of gravity among the three DNS cases.  For each $Fr$, particles of six Stokes numbers 
in the range $ 0.01 \le St_\eta \le 0.2$ were tracked in the DNS, and particle 
clustering quantified both as a function of separation and the spherical polar angle.   
We compared the DNS and theory values for the 
power-law exponent $\beta$ characterizing the dependence of clustering on 
separation.  Reasonable agreement is seen between 
the DNS $\beta$'s for the $Fr = 0.006$ case and the theoretical predictions obtained 
using the second drift closure (referred to as DF2).  Further, in conformity with  
the DNS, theory shows that the clustering of $St_\eta \ll 1$ particles 
is only weakly anisotropic.

\end{abstract}

\section{Introduction}
This paper presents the Part II of the current work on a stochastic theory for  
the clustering of monodisperse, low-inertia particle pairs that are settling rapidly 
in homogeneous isotropic turbulence.  The theory is developed in the limits of 
$Fr \ll St_\eta \ll 1$, where the Stokes number $St_\eta$ is the ratio of the 
particle response time to the Kolmogorov time scale. 
The Part I paper presented the derivation of: (1) closure approximations for the drift and 
diffusion fluxes in the probability density function (PDF) equation for  
pair relative positions, and (2) the analytical solution to the 
PDF $\langle P \rangle (r,\theta)$, where $r$ is the separation, and $\theta$ is the spherical polar angle.  
Part II focuses on the quantitative 
analysis of the theory by comparing its predictions of particle clustering with the 
results from DNS of isotropic turbulence containing particles settling under gravity.

Gravitational settling modifies particle dynamics in important ways.  
One of the main effects of gravity is that it introduces anisotropy 
into the particle sampling of the underlying turbulence, and thereby 
in the spatial clustering of particles.  Gravity also 
alters the correlation times of fluid velocity gradients along particle trajectories.  The 
altered time scales, in turn, modulate the path-history effects that play 
a key role in determining particle clustering (Part I paper presents a 
more detailed discussion of the path-history effects and their role in clustering).  
The theory developed in this study incorporates the effects on clustering of 
settling-induced anisotropy, as well as of settling-modulated flow time scales.   

Recently, \citet{peter2015b} performed a detailed DNS investigation of the 
effects of gravity on the dynamics of single particles, as well as particle 
pairs.  In their study, the Froude number $Fr = 0.052$, which is representative of 
fluid accelerations in cumulus clouds. They considered a wide range of Taylor 
micro-scale Reynolds numbers ($88 \leq$ \Rel $\leq 597$), 
Stokes numbers ($0 \leq St_\eta \leq 56.2$).   
For $St_\eta < 1$, they showed that the principal effect of gravity 
on particle clustering was to decrease the inward (radial) drift, thereby 
reducing the radial distribution function (RDF).
They also found that gravity mitigates the preferential concentration mechanism by reducing  
the interaction time between particles and the underlying turbulence.  
Specifically, gravity reduces the Lagrangian time scales of strain-rate and 
rotation-rate along particle trajectories.  As shown in \cite{chun2005}, the drift flux 
is proportional to the time integral of the two-time correlation 
of $[S^2(t) - R^2(t)]$ along the trajectory of the primary particle, 
where $S^2$ and $R^2$ are the second invariants 
of the strain-rate and rotation-rate tensors, respectively.   
\citet{peter2015b} also quantified the anisotropy in particle clustering due to gravity through 
the use of spherical harmonic functions to represent the RDF dependence on the polar 
angle $\theta$.

\citet{bec2014gravity} performed a DNS study of the effects of settling 
on inertial particle clustering in isotropic turbulence.  They considered both 
low and high Stokes number particles ($St_\eta \lesssim 1$ and $St_\eta > 1$, respectively) 
that are settling under low- and high-gravity conditions ($Fr > 1$ and $Fr < 1$, respectively).
\citet{bec2014gravity} observed that when $Fr \ll 1$, the clustering of $St_\eta \lesssim 1$ 
particles decreased and became anisotropic in that particles formed streaks 
along the vertical (or gravity) direction.  The opposite effect is seen for  
$St_\eta > 1$ particles, whose clustering increased when compared to the zero-gravity case.  
The opposing effects of gravity on the clustering of 
low- and high-Stokes number particles was also seen by \citet{peter2015b}.  
Other DNS studies of settling particles 
\citep{ayala08a,onishi2009,woittiez2009,parishani2015} focused primarily on the 
collision rates of droplets in isotropic turbulence.

In this Part II paper, we present a comparison of the theory-predicted particle clustering  
with the corresponding DNS data.  For Stokes numbers $St_\eta \ll 1$, we compare 
the exponent $\beta$ characterizing the power-law dependence 
of the PDF $\langle P \rangle (r,\theta)$ on separation $r$.  We also compare 
the degree of anisotropy of clustering obtained from the theory with that 
from DNS.  In the case of DNS, 
particle clustering is quantified at $Re_\lambda = 77.76$ for three Froude numbers 
$Fr = \infty,~ 0.052,~ 0.006$ (in the order of increasing gravity), and 
six Stokes numbers in the range $0.01 \le St_\eta \le 0.2$.  For the highest 
gravity case, $Fr = 0.006$, the particle settling time through the periodic box 
of length $2\pi$ is close to the integral time scale.  This may lead to errors 
since particles that have exited the domain and been reintroduced into it will 
again encounter the same correlated eddies that they have already seen  
previously on their way down the box \citep{peter2015b,woittiez2009}. 
To eliminate this numerical artifact, we performed 
DNS with a bigger domain size of $4\pi$ along the vertical direction for the $Fr = 0.006$ case.

The organization of the paper is as follows. Section \ref{sec:sims} presents 
the computational details of the DNS runs, as well as the 
particle evolution algorithm.   Quantification of the two-time correlations of 
dissipation rate and enstrophy is discussed in Section \ref{sec:ll}.  The development 
of model energy spectrum that closely matches the DNS energy spectrum is 
presented in Section \ref{sec:inputs}.  
In section \ref{sec:results}, we present the comparison 
of theory predictions of particle clustering with the DNS data. 
Section \ref{sec:conclusions} summarizes the key 
findings of this Part II paper.  

\section{Computational Method}
\label{sec:sims}
\subsection{Fluid Phase}
\label{subsec:gov_fluid}
Direct numerical simulations of forced isotropic turbulence 
were performed using a pseudo-spectral method based on the discrete 
Fourier expansions of flow variables. The simulation domain, consisting 
of a cube of length $2\pi$, is discretized into $N^3$ grid points, with 
periodic boundary conditions along the three cartesian directions.

The govening equations for the flow are the Navier-Stokes equations in  
rotational form and the continuity equation \citep{ireland13,brucker07}
\begin{gather}
\frac{\partial \bs{u}}{\partial t} + {\boldsymbol{\omega}} \times \bs{u} =
-\bs{\nabla}\left ( p/\rho_f + \bs{u}^2/2  \right ) + \nu \bs{\nabla}^2 \bs{u} 
 \label{eq:mom} \\ 
\bs{\nabla} \bcdot \bs{u} = 0 \label{eq:cont}
\end{gather}
where $\bs{\omega} = \bs{\nabla} \times \bs{u}$ is the vorticity, $\rho_f$ is the fluid
density, and $p$ is the pressure.

Transforming Eqs.~\eqref{eq:mom} and \eqref{eq:cont} into Fourier space and 
eliminating pressure using the spectral form of continuity yields  
\begin{gather}
\label{eq:fourierns}
\left ( \frac{\partial}{\partial t} + \nu k^2  \right ) \bs{\widehat{u}} = 
-\left ( \bs{I} - \frac{\bs{k} \bs{k}} {k^2} \right ) \bcdot 
\widehat{ {\boldsymbol{\omega}} \times \bs{u} } 
\end{gather}
where $k^2 = \bs{k} \bcdot \bs{k}$. 
Direct evaluation of the convolution $\widehat{ {\boldsymbol{\omega}} \times \bs{u} }$ 
is extremely computationally intensive.  Hence, a pseudo-spectral approach is adopted wherein 
${ {\boldsymbol{\omega}} \times \bs{u} }$ is first computed in physical space, and then 
transformed into the spectral space. 

Since the time-derivative and viscous stress terms on the LHS of Eq.~\eqref{eq:fourierns} 
are linear in $\bs{\widehat{u}}$, one may evolve these terms in time exactly  
by multiplying Eq.~\eqref{eq:fourierns} with the integrating factor, $\text{exp}(\nu k^2t)$.  This 
yields the following equation (in index notation):
\begin{equation}
\frac{\partial }{\partial t} \big [\exp \big(\nu k^2t \big )~\widehat{u}_i \big ]  = 
\text{RHS}_i~\exp \big(\nu k^2t \big), \label{temp_dis}
\end{equation}
where $\text{RHS}_i = \left(-\delta_{im} + \frac{k_ik_m}{k^2}\right)\epsilon_{mjk}\mathscr{F}\{\omega_j u_k\}$ 
represents the right-hand side of Eq.~\eqref{eq:fourierns}, and $\epsilon_{mjk}\mathscr{F}\{\omega_j u_k\}$ 
represents the convolution $\widehat{ {\boldsymbol{\omega}} \times \bs{u} }$, and 
$\epsilon_{mjk}$ is the Levi-Civita tensor. 

Equation \eqref{temp_dis} is then discretized in time using the second-order Runge-Kutta (RK2) method giving  
\begin{equation}
\widehat{u_i}^{n+1} = \widehat{u_i}^n~\text{exp} \big(-\nu k^2t \big) + 
\frac{}{} \big \{\text{RHS}_i^n~\text{exp} \big(-\nu k^2t \big) + \text{RHS}_i^{n+1} \big \}
\end{equation}
where $n$ is the the previous time-step level and $h$ is the time-step size. To prevent convective instabilities, time-step size 
$h$ is chosen such that the CFL number $\leq 0.5$.
The pseudospectral algorithm introduces aliasing errors which are removed by zeroing the fluid velocities in spectral space for wavenumbers satisfying $k \geq k_{\rm max}$, where $k$ is the wavenumber magnitude, 
$k_{\rm max} = \sqrt{2}{\text N}/3$, and $N$ is the number of grid points along each dimension.

To achieve statistically stationary turbulence, we employ the deterministic 
forcing method developed by \cite{witkowska1997}, wherein 
the turbulent kinetic energy dissipated during a time step is added back to the 
flow at the low wavenumbers.  It may be noted that in this method, there is no explicit forcing 
term $\bs{f}$ added to the Navier-Stokes equations.  Instead, one  scales the velocity
components in the wavenumber band [$\kappa_{\rm min},\kappa_{\rm max} $] 
by a factor such that the energy dissipated during a given time step is 
resupplied, as follows.
\begin{equation}
\widehat{\bs{u}}(\bs{\kappa},t+\Delta t) = \widehat{\bs{u}}(\bs{\kappa},t+\Delta t)\sqrt{1+\frac{\Delta E_{\rm diss}(\Delta t)}{\int_{\kappa_{\rm min}}^{\kappa_{\rm min}} E(\kappa, t+\Delta t)d\kappa}} ~~~\forall ~\kappa \in [\kappa_{\rm min},\kappa_{\rm max} ]
\label{eq:brass_det}
\end{equation}
where $\widehat{\bs{u}}(\bs{\kappa},t)$ is the 
spectral velocity, $\Delta E_{\rm diss}$ is the total energy dissipated 
during $\Delta t$, and $E(\kappa, t+\Delta t)$ is the spectral turbulent kinetic energy 
in a wavenumber shell with magnitude $\kappa$ at time $t+\Delta t$.
In the current study, the velocity components in the range $\kappa \in (0,\sqrt{2}]$ are forced 
using Eq.~\eqref{eq:brass_det}.

\subsection{Particle Phase}
\label{subsec:gov_particle}

The governing equations of motion for a heavy spherical particle, whose diameter 
is much smaller than the Kolmogorov length scale, may be written as
\begin{align}
\frac{d\bs{x}_p}{dt} & = \bs{v}_p, \label{eq:part1}\\
\frac{d\bs{v}_p}{dt} & = \frac{\bs{u}(\bs{x}_p,t)-\bs{v}_p}{\tau_v} 
+ \bs{g}, \label{eq:part2}
\end{align}
where we assumed Stokes drag to be the principal force on the particle, 
$\bs{x}_p$ and $\bs{v}_p$ are the particle position and velocity, respectively, and 
$\tau_v$ is the particle viscous relaxation time.  
In Eq.~\eqref{eq:part2}, $\bs{u}(\bs{x}_p,t)$ 
is the fluid velocity at the particle's location.  
We neglect two-way coupling effects, as well as particle collisions. 
In order to solve Eqs.~\eqref{eq:part1} and \eqref{eq:part2} numerically, 
$\bs{u}(\bs{x}_p,t)$ 
needs to be evaluated. This is achieved by interpolating, to the particle 
position, fluid velocities at a 
stencil of grid points surrounding the particle.  We use the 
8$^\text{th}$ order Lagrange interpolation method that is based on a stencil of 
$8 \times 8 \times 8$ fluid velocities.

Temporal update of particle motion is achieved through a modified 
second-order Runge-Kutta (RK2) method in which the standard RK2 weights 
are replaced by exponential integrators as follows \cite{ireland13}.
\begin{equation}
\bs{v}_p(t_0+h) = e^{-h/\tau_v}~{\bs{v}_p}(t_0) + w_1~{\bs{u}}[\bs{x}_p(t_0)] + 
w_2~{\bs{u}}[\bs{x}_p(t_0) + \bs{v}_p(t_0)h] + (1-e^{-h/\tau_v})\tau_v\bs{g},
\end{equation}
where $h$ is the time step, and the exponential integrators $w_1$ and $w_2$ are given by 
\begin{equation}
w_1 \equiv \left(\frac{h}{\tau_v}\right) \left[\phi_1 \left(\frac{-h}{\tau_v}\right) - \phi_2 \left(\frac{-h}{\tau_v}\right)\right], ~ w_2 \equiv \left(\frac{h}{\tau_v}\right)\phi_1 \left(\frac{-h}{\tau_v}\right)
\end{equation}

\begin{equation}
\phi_1(z) \equiv \frac{e^z-1}{z}, ~ \phi_2(z) \equiv \frac{e^z-z-1}{z^2}
\end{equation}

For small values of $Fr$, the periodic box length $L$ is an important consideration 
since it can artificially influence the motion of settling, inertial particles.  
Specifically, the use of periodic boundary conditions is problematic if the time it takes the settling  
particles to traverse the length $L$ is $\leq O(T_E)$, where $T_E$ is the large eddy turnover time. 
Several studies \citep{woittiez2009,peter2015b,dhariwal18jfm}  considered this issue in detail and found that 
box sizes larger than $L = 2\pi$ may be needed, particularly when $St_\eta \gtrsim 1$. 
In this paper, the smallest Froude number is $Fr = 0.006$.   
Considering $St_\eta = 0.1$, $Fr = 0.006$, and $T_E = 1.568$, it can be shown that the settling 
times of particles through a domain length $L = 2\pi$ is
$T_{\rm settle}^{L = 2 \pi} \approx 1.75$. 
Since $T_{\rm settle}^{L = 2 \pi}$ is only margninally greater than 
$T_E$, we considered a 
box dimension of $4\pi$ along the direction of gravity for the $Fr = 0.006$ case.

\section{Correlation Times in Second Drift Closure: $T_{\epsilon \epsilon}$, $T_{\zeta \zeta}$, 
$T_{\epsilon \zeta}$, $T_{\zeta \epsilon}$}
\label{sec:ll}
The drift flux $q_i^d(\boldsymbol{r},t)$ in the transport equation 
for the PDF $\langle P \rangle$ is given by  
\begin{gather}
q_i^d(\boldsymbol{r},t) = -\langle P\rangle(\boldsymbol{r};t)~\frac{St_\eta^2}{\Gamma_\eta^2}~ r_k 
\int_{-\infty}^{t} d_{ik}~dt' \label{eq:drift_flux_gamma_terms}
\end{gather}
where $\Gamma_\eta$ is the inverse of the Kolmogorov time scale.  
In the Part I paper, we derived two closure forms for the integral on the 
right-hand side (RHS) of \eqref{eq:drift_flux_gamma_terms}, referred to as DF1 and DF2.  
DF1 is based on the assumption that the fluid velocity gradient along particle trajectories 
has a Gaussian distribution.  In DF2, we regard the strain-rate and rotation-rate tensors 
scaled by the turbulent dissipation rate and enstrophy, respectively, as  
normally distributed.  

In the second closure of drift flux (DF2), this integral is given by
\begin{gather}
\int_{-\infty}^{t} d_{ik}~dt' = 
\frac{1}{4\nu^2} \Bigl \{ ~\frac{1}{3} \delta_{ik}
\left [ \langle \epsilon^2 \rangle T_{\epsilon \epsilon} + 
\langle \epsilon \zeta \rangle T_{\epsilon \zeta} - 
\langle \zeta \epsilon \rangle T_{\zeta \epsilon} - 
\langle \zeta^2 \rangle T_{\zeta \zeta} 
\right ] +  \notag \\
2 \langle \epsilon^2 \rangle \int_{-\infty}^{t} 
\exp \left ( -\frac{t-t'}{T_{\epsilon \epsilon} } \right ) \langle \sigma_{ij}(t) ~\sigma_{lm}(t') \rangle ~
\langle \sigma_{jk}(t) ~\sigma_{lm}(t') \rangle~dt' - \notag \\
2 \langle \zeta^2 \rangle \int_{-\infty}^{t} 
\exp \left ( -\frac{t-t'}{T_{\zeta \zeta} } \right ) \langle \rho_{ij}(t) ~\rho_{lm}(t') \rangle ~
\langle \rho_{jk}(t) ~\rho_{lm}(t') \rangle~dt'
~\Bigr \}.
\label{eq:int_rhs_dik}
\end{gather}
Equation \eqref{eq:int_rhs_dik} contains the auto- and cross-correlation 
times of the dissipation rate $\epsilon$ and enstrophy $\zeta$---$T_{\epsilon \epsilon}$, $T_{\zeta \zeta}$, $T_{\epsilon \zeta}$, and $T_{\zeta \epsilon}$.  These are approximated to be along 
the trajectories of fluid particles collocated with the inertial particles.  Furthermore, 
since the particles are settling rapidly ($Sv_\eta = g\tau_v/u_\eta \gg 1$), we may 
regard the surrounding turbulence as essentially 
frozen during a particle response time $\tau_v$.  Therefore, based on the 
Taylor's hypothesis, the Lagrangian time scales may be 
expressed as the respective spatial correlation lengths divided by the particle settling velocity. 
For instance, $T_{\epsilon \epsilon} = L_{\epsilon \epsilon}/g\tau_v$, and 
$T_{\epsilon \zeta} = L_{\epsilon \zeta}/g\tau_v$.  Due to isotropy, 
$L_{\epsilon \zeta} = L_{\zeta \epsilon}$, so that the terms $\langle \epsilon \zeta \rangle T_{\epsilon \zeta}$ 
and $(-\langle \zeta \epsilon \rangle T_{\zeta \epsilon})$ cancel out 
on the RHS of \eqref{eq:int_rhs_dik}.  Therefore, the unknown length scales are 
$L_{\epsilon \epsilon}$ and $L_{\zeta \zeta}$.  
We now discuss the procedure for computing the 
length scale $L_{\epsilon \epsilon}$ through DNS (an analogous process 
is used to compute $L_{\zeta \zeta}$).    

The length scale $L_{\epsilon \epsilon}$ is defined as
\begin{eqnarray}
L_{\epsilon \epsilon} = \frac{\int R_{\epsilon \epsilon}(r) dr }{\langle \epsilon^2 \rangle}
\end{eqnarray}
where $R_{\epsilon \epsilon}(r) = 
\langle \epsilon(\bs{x};t) ~ \epsilon(\bs{x}+\bs{r};t)\rangle$ is the spatial correlation of 
dissipation rate.  It is evaluated using Fourier transforms as
\begin{eqnarray}
R_{\epsilon \epsilon}(r) = \int d{\bs{\kappa}} ~ \Phi_{\epsilon \epsilon}(\bs{\kappa})~
e^{\mi  \bs{\kappa} \bcdot \bs{r}}
\end{eqnarray}
where $\Phi_{\epsilon \epsilon}$ is the Fourier coefficient of $R_{\epsilon \epsilon}$, and 
$\bs{\kappa}$ and $\bs{r}$ are the wavenumber and relative position vectors, respectively.
Expressing the integral $\int d{\bs{\kappa}}$ in spherical coordinates, and performing 
the integrations in polar and azimuthal angles, we have
\begin{eqnarray}
R_{\epsilon \epsilon}(r) = \int d\kappa ~D_\epsilon(\kappa)~
\frac{\sin(\kappa r)}{\kappa r} \label{eq:int_wavenummag}
\end{eqnarray}
where $\kappa = |\bs{\kappa}|$ and 
\begin{eqnarray}
D_\epsilon(\kappa) = \left \langle \sum_{|\bs{\kappa}| = \kappa} \widehat{\epsilon}(\bs{\kappa},t) ~
\widehat{\epsilon}^*(\bs{\kappa},t) \right \rangle
\label{eq:d_eps_averaged}
\end{eqnarray}
In the above equation, $\widehat{(\cdot)}$ denotes the Fourier coefficient, and the 
superscript $*$ the complex conjugate. 
We evaluate $D_\epsilon(\kappa)$ using DNS, while the integral in \eqref{eq:int_wavenummag} 
is calculated through numerical quadrature.  In \eqref{eq:d_eps_averaged}, 
$\langle \cdots \rangle$ denotes averaging over an ensemble of temporal snapshots 
of the statistically stationary turbulent velocity field.  The length scales thus 
determined from the current DNS are given in Table \ref{tab:eps_zeta_scales}.
\begin{table}
\centering
\renewcommand{\arraystretch}{1.1}
\setlength{\tabcolsep}{32pt}
  \begin{tabular}{cc}
    $L_{\epsilon \epsilon}$ & $L_{\zeta \zeta}$  \\ \hline
    0.257 & 0.211  \\ \hline
  \end{tabular}
  \caption{Correlation length scales of $\epsilon$ and $\zeta$ for $Re_\lambda = 77.76$.}
  \label{tab:eps_zeta_scales}
\end{table}

\section{DNS Inputs to Theory}
\label{sec:inputs}
\begin{table}
\centering
\begin{tabular}{|c|c|}
\hline
Parameter                & DNS I     \\ \hline
$N$                      & 128      \\
$Re_\lambda$             & 77.756   \\
$u_{\rm rms}$            & 0.968      \\
$\nu$                    & 0.0071     \\
$\epsilon$               & 0.307      \\
$L$                      & 1.518     \\
$\lambda$                & 0.572     \\
$\eta$                   & 0.033   \\
$T_E$                    & 1.568     \\
$\tau_\eta$              & 0.153     \\
$\kappa_{{\rm max}}\eta$ & 1.991     \\
$\Delta t$               & $2.5\times 10^{-3}$  \\
$N_p$                    & 300,000  \\
\hline
\end{tabular}
\caption{Simulation parameters for the DNS study. All dimensional parameters are in arbitrary units.
          $Re_\lambda \equiv u_{\rm rms}\lambda/\nu$ is the Taylor micro-scale
          Reynolds number, $u_{\rm rms} \equiv \sqrt{(2k/3)}$ is the fluid RMS fluctuating 
          velocity, $k$ is the turbulent kinetic energy, 
          $\nu$ is the fluid kinematic viscosity,
          $\epsilon \equiv 2\nu \int_0^{\kappa_{\rm max}}\kappa^2 E(\kappa) ~{\rm d}\kappa $ is the
          dissipation rate of turbulent kinetic energy, 
          $L \equiv 3\pi/(2k)\int_0^{\kappa_{\rm max}}E(\kappa)/\kappa ~{\rm d}\kappa $
          is the integral length scale, $\lambda \equiv 
          u_{\rm rms}\sqrt{(15\nu/\epsilon)}$ is the Taylor microscale,
          $\eta \equiv \nu^{3/4}/\epsilon^{1/4}$ is the Kolmogorov length scale, 
          $T_E \equiv L/u_{\rm rms}$ is the large-eddy turnover
          time, $\tau_\eta \equiv \sqrt{(\nu/\epsilon)}$ is the Kolmogorov time scale, 
          $\kappa_{\rm max}$ is the maximum
          resolved wavenumber, $\Delta t$ is the time step, and $N_p$ is  
          the number of particles per Stokes number.}
\label{tab:128_parameters}
\end{table}

To be able to consistently compare theory and DNS, it is 
important that the theory use the same turbulence parameters as those 
in statistically stationary DNS.  Hence, inputs to the theory such as the 
Kolmogorov and integral length scales, dissipation rate, kinematic viscosity, 
root-mean-square velocity $u_{\rm rms}$, and 
$\Rey_\lambda$ are all identical to those in Table~\ref{tab:128_parameters}, which 
lists the DNS turbulence parameters.  
In particular, one also has to ensure that the energy spectrum 
$E(\kappa)$ needed in the theory (to compute the drift and diffusion flux 
coefficients) closely matches the DNS energy spectrum.  
This was achieved by suitably selecting the parametric inputs to the model spectrum 
provided in \citet{pope_2000}, as follows:

\begin{eqnarray}
E(\kappa) &=& C\epsilon^{2/3}\kappa^{-5/3}f_L(\kappa L)f_\eta(\kappa \eta) \label{eq:modelenergy}\\
f_L(\kappa L) &=& \left(\frac{\kappa L}{[(\kappa L)^2+c_L]^{1/2}}\right)^{5/3+p_0} \label{eq:fl} \\
f_\eta(\kappa \eta) &=& \mathrm{exp}\left \{-\beta\left ( [(\kappa \eta)^4 + c_\eta^4]^{1/4} -c_\eta\right ) \right \} \label{eq:feta}
\end{eqnarray}
where $\beta = 5.2$ and $p_0 = 2$ \citep{pope_2000}.
\begin{table}
\centering
\renewcommand{\arraystretch}{1.1}
\setlength{\tabcolsep}{32pt}
  \begin{tabular}{cc}
    $\mathrm{Parameter}$ & Value \\ \hline
       $C$ & 1.908  \\ 
       $c_L$ & 0.2855 \\
       $c_\eta$ & 0.2165  \\ 
       $\kappa_{{\rm max}}$ & 60 \\ \hline
  \end{tabular}
  \caption{Parameters for the model energy spectrum at $\Rey_\lambda$ =  77.76.  
  After determining $c_L$ and $c_\eta$, 
  the parameter $C$ was adjusted to match the DNS energy spectrum. \cite{pope_2000} 
  suggested $C = 1.5$.}
  {\label{tab:modelparameters}}
\end{table}
\begin{figure}
  \begin{center}
   \psfrag{x}[cc][2]{\Large{$\kappa$}}
   \psfrag{b}[cc][2]{\Large{${E(\kappa)}$ }}
   \psfrag{p}[cc][2]{\small{ DNS}}
   \psfrag{r}[cc][2]{\small{ Model}}

   \includegraphics[scale=0.4]{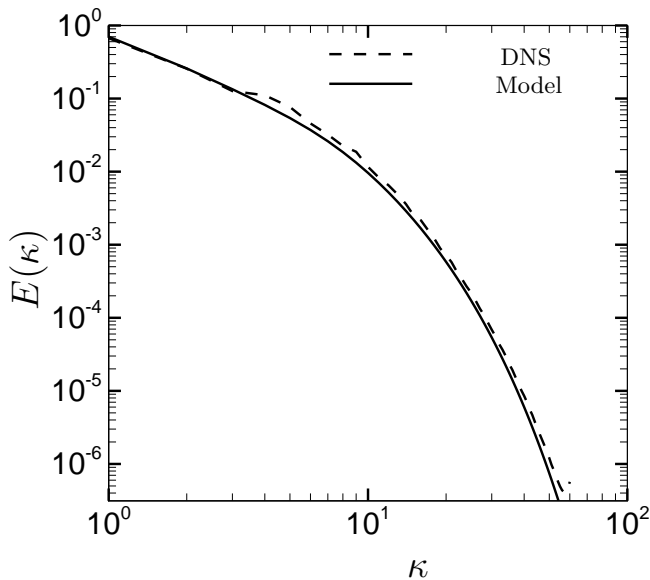}
  \end{center}
  \caption{Comparison of the DNS and model energy spectra at $Re_\lambda = 77.76$.}
  \label{fig:energyspectra}
\end{figure}
The parameters $ c_L$ and $c_\eta $ are determined from the following constraints:
\begin{gather}
\frac{3}{2} u_{\rm rms}^2  =  \int_1^{\kappa_{\rm max}} E(\kappa) {\rm d}\kappa \label{eq:kenergy}\\
\epsilon  =  2\nu \int_1^{\kappa_{\rm max}}\kappa^2 E(\kappa) {\rm d}\kappa \label{eq:epsilon}
\end{gather}
where $\epsilon$ is the dissipation rate, and the wavenumber limits $[1,\kappa_{\rm max}]$ 
are the same as in DNS. These wavenumber limits are also used in 
calculating the drift and diffusion flux coefficients.  The parameters $c_L$ and $c_\eta$ 
are numerically evaluated using the DNS values of $u_{\rm rms}$, $\epsilon$  and $\nu$ 
from Table~\ref{tab:128_parameters}.  The resulting values of $c_L$ and $c_\eta$ are shown in 
Table~\ref{tab:modelparameters}.  In figure~\ref{fig:energyspectra}, the model spectra 
calculated from equations~\eqref{eq:modelenergy}-\eqref{eq:feta} are compared with 
the DNS energy spectrum for $\Rey_\lambda$ =  77.76.  Good agreement is seen 
between the model and DNS spectra.

\section{Results}
\label{sec:results}
We have shown in the Part I paper that the PDF of pair separation is given by 
\begin{eqnarray}
\langle P \rangle(r,\mu)=r^{\beta_2 St_\eta^2} \frac{1}{4\pi}\left[1+St_\eta^2 ~\beta_2  
\left (\frac{1}{2} \ln(1-\mu^2) - (\ln 2 - 1) \right ) \right]
\label{eq:pdf_sol}
\end{eqnarray}
where $\mu = \cos \theta$, with $\theta \in (-\pi,\pi)$ being the spherical polar angle. 
The power-law exponent $\beta_2$ is given by
\begin{eqnarray}
\beta_2=\frac{\lambda_2+2\lambda_1}{\alpha_2}
\label{eq:solmthd6}
\end{eqnarray}
where $\lambda_1$ and $\lambda_2$ are the drift flux coefficients, and $\alpha_2$ is the 
diffusion flux coefficient.  These coefficients are determined through a numerical 
quadrature process, discussed in Part I.  
It may be recalled that there are two forms of $\lambda_1$ and $\lambda_2$, corresponding to 
the two drift closures DF1 and DF2.  Thus, we will compare two theoretical values of $\beta_2$ 
with the corresponding DNS value.

DNS were performed at a Taylor micro-scale Reynolds number $Re_\lambda = 77.76$ 
for three Froude numbers $Fr = \infty,~ 0.052,~ 0.006$, where 
$Fr = \infty$ corresponds to zero gravity, and $Fr = 0.006$ to the highest 
magnitude of gravity considered in the current DNS.  For each $Fr$, particles of six Stokes numbers 
$St_\eta = 0.01,~0.02,~ 0.04,~ 0.08,~ 0.15,~ 0.2$ were tracked.  Spatial clustering of particles 
is quantified through the radial distribution function (RDF), $g(r)$, which scales 
with separation $r$ as 
$g(r) \sim r^{\beta_2 St_\eta^2}$ for $r \ll \eta$.  The DNS value of the exponent $\beta_2$ is then 
determined through a least squares curve fit of the RDF for $0.6 \eta \le r \le 2.5 \eta$.

\begin{figure}

    \centering

        \psfrag{a}{\Large{$St_\eta$}}
        \psfrag{b}[c]{\Large{$\beta = -\beta_2 \times St_\eta^2$}}
        \psfrag{c}{$\infty$}
        \centering
        \hspace{-0.0in}
        \includegraphics[scale = 0.5]{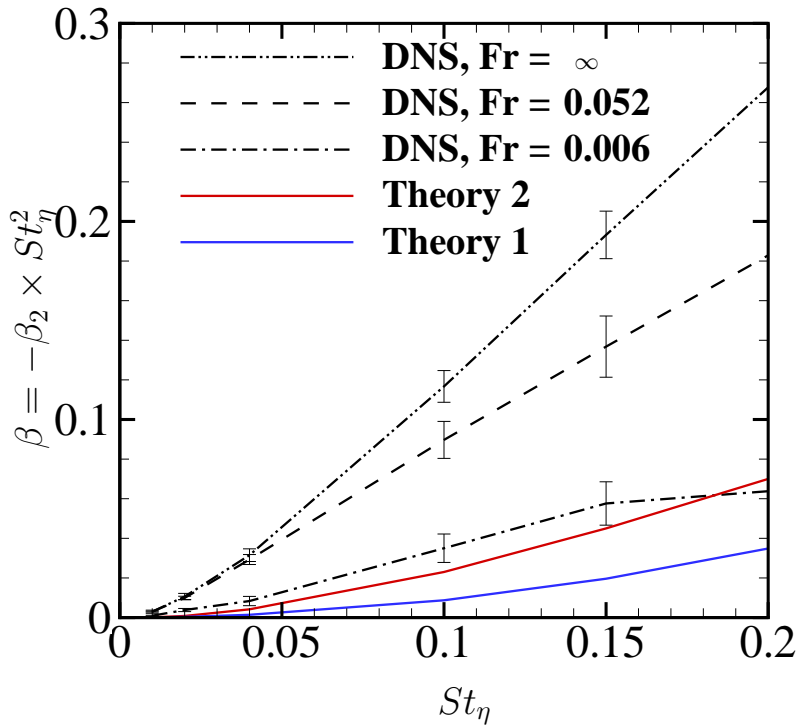}
    \caption{Comparison of the power-law exponent $\beta$ obtained from theory and DNS. 
    Results obtained using both DF1 and DF2 are shown (referred to as 
    Theory~1 and Theory~2, respectively). There is an uncertainty of 
 $\sim 8$\% in the DNS values of $\beta$.}
    \label{fig:beta2_comp}
\end{figure}

In figure \ref{fig:beta2_comp}, we compare the exponent $(-\beta_2 \times St_\eta^2)$ obtained from 
DNS and theory.  The theoretical $\beta_2$'s for both 
DF1 and DF2 (valid in the $Fr \ll 1$ limit), and the DNS $\beta_2$'s for the three $Fr$'s are presented in 
figure \ref{fig:beta2_comp}.  A key feature of DF2 is that it 
accounts for the two-time autocorrelations and cross-correlations of dissipation rate 
and enstrophy (or, equivalently of the second invariants of strain-rate and 
rotation-rate tensors).  \cite{chun2005} showed that these correlations quantify, as well as illustrate  
the mechanisms driving the clustering of non-settling particles. 
It is evident from figure \ref{fig:beta2_comp} that the DF1 $\beta_2$'s are 
significantly lower than those of DNS for all three Froude numbers.
However, the DF2 $\beta_2$'s are 
in reasonable agreement with the DNS $\beta_2$'s for $Fr = 0.006$.  The improved 
performance of DF2, compared to DF1, is because DF2 accounts 
for the correlations of dissipation rate and enstrophy, but DF1 does not.

Next, we consider the anisotropy in particle clustering due to settling.  
\citet{peter2015b} quantified the anisotropy through the use of the 
angular distribution function (ADF), $g(\bs{r})$, and expressed 
it in terms of the Legendre spherical harmonic functions, as below.
\begin{equation}
\frac{g(\bs{r})}{g(r)} = \sum_{l = 1}^{\infty} \frac{\mathscr{C}_{2l}^0(r)}{\mathscr{C}_0^0(r)}
Y_{2l}^0(\cos\theta)
\label{eq:gr_ratio}
\end{equation}
where 
\begin{equation}
g(r) = \mathscr{C}_0^0(r) = \int_{0}^{\pi} d\theta~ \sin\theta~ g(\bs{r})
\end{equation}
describes the dependence of clustering on separation $r$. 
Applying the orthogonality of Legendre polynomials to \eqref{eq:gr_ratio}, we get
\begin{eqnarray}
\frac{\mathscr{C}_{2}^0(r)}{\mathscr{C}_0^0(r)} = \frac{5}{2} 
\frac{\int_{0}^{\pi} d\theta~ \sin\theta~ g(\bs{r})~Y_{2}^0(\cos\theta)} {g(r)} 
\end{eqnarray}
The theoretical value of the coefficient ratio is 
\begin{eqnarray}
\left [\frac{\mathscr{C}_{2}^0(r)}{\mathscr{C}_0^0(r)} \right ]_{\rm theory} &=& \frac{5}{2} 
\frac{\int_{0}^{\pi} d\theta~ \sin\theta~ \langle P \rangle(r,\theta)~Y_{2}^0(\cos\theta)} 
{\int_{0}^{\pi} d\theta~ \sin\theta~ \langle P \rangle(r,\theta)} \notag \\
&=& \frac{5\beta_2 St_\eta^2}{12}
\label{eq:c2ratio_theory}
\end{eqnarray}
In table \ref{tab:coeff}, we compare the values of ${\mathscr{C}_{2}^0(r)}/{\mathscr{C}_0^0(r)}$ 
obtained using theory and DNS, the latter for $Fr = 0.006$.  We see that the degree of anisotropy 
predicted by the theory is in reasonable agreement with that computed using DNS, particularly 
for $St_\eta \le 0.10$.  However, the theory generally overpredicts the coefficient ratio 
as compared to DNS.

\begin{table}
\centering
\begin{tabular}{|c|c|c|}
\hline
$St_\eta$   &  Theory (DF2)  &  DNS ($Fr = 0.006$)  \\ \hline
0.01 & $4.17 \times 10^{-5}$ & $3.84 \times 10^{-4}$  \\
0.02 & $3.33 \times 10^{-4}$ & $2.82 \times 10^{-4}$  \\
0.04 & $1.73 \times 10^{-3}$ &  $1.42 \times 10^{-3}$ \\
0.05 & $2.50 \times 10^{-3}$ & $2.23 \times 10^{-3}$  \\
0.10 & $9.58 \times 10^{-3}$ &  $8.68 \times 10^{-3}$ \\
0.15 & $1.88 \times 10^{-2}$ &  $1.12 \times 10^{-2}$ \\
0.20 & $2.92 \times 10^{-2}$ & $1.42 \times 10^{-2}$ \\
\hline
\end{tabular}
\caption{Comparison of ${\mathscr{C}_{2}^0(r)}/{\mathscr{C}_0^0(r)}$ obtained using 
theory (DF2 only) and current DNS for $Fr = 0.006$.  There is an uncertainty of 
 $\sim 10$\% in the DNS values of the coefficient ratio.}
\label{tab:coeff}
\end{table}

\section{Conclusions} 
\label{sec:conclusions}
Part II of this study focuses on the quantitative analysis of the 
theory through a direct comparison of theory predictions 
with the data obtained in our DNS runs.  While the theory is derived 
in the $Fr \ll 1$ regime, 
DNS were performed for three Froude numbers $Fr = \infty,~ 0.052,~ 0.006$ 
at $Re_\lambda = 77.76$.  In the DNS runs, the $Fr = 0.006$ case 
represented the run with the highest magnitude of gravitational acceleration.  For this 
case, a domain size of $4 \pi$ was used in the direction of gravity 
to mitigate the numerical errors arising from particles spuriously sampling the 
same eddies more than once as they settle through the domain.  A model energy spectrum 
was derived that closely matched the DNS energy spectrum.  
The correlation length scales of dissipation rate and enstropy, $L_{\epsilon \epsilon}$ 
and $L_{\zeta \zeta}$, needed for DF2 were also obtained using DNS.  
The model energy spectrum, and the two correlation lengths were  
then used in the calculation of the drift and diffusion coefficients needed for the 
power law exponent, as well as the coefficient ratio (for quantifying anisotropy). 
In terms of the dependence of clustering on separation, we see that the DF1 $\beta$'s 
are significantly lower than the DNS $\beta$'s for the three Froude numbers.  But, the DF2 $\beta$'s are in 
reasoanble quantitative agreement with the DNS values for $Fr = 0.006$.  We also 
quantified the anisotropy in particle clustering due to the settling.  We 
see that the values of the coefficient ratio ${\mathscr{C}_{2}^0(r)}/{\mathscr{C}_0^0(r)}$ obtained 
using DF2 are in reasonable agreement with the DNS values at $Fr = 0.006$, particularly for 
$St_\eta \le 0.1$.  These results demonstrate that the the two-time 
correlations of dissipation rate and enstropy constitute an important mechanism 
driving the drift flux responsible for clustering.  The current two-part study 
presents the development and analysis of an analytical theory for the clustering 
of low-inertia particle pairs that are settling rapidly in isotropic turbulence.  
We see that the theory accurately captures the quantitative trends in particle 
clustering.  It also shows reasonable quantitative agreement with DNS data 
for low Stokes and Froude numbers.

\section*{Acknowledgements}

SLR gratefully acknowledges NSF support through the grant CBET-1436100.

\bibliographystyle{jfm}
\bibliography{refpaper2}

\end{document}